\documentclass[twocolumn,twoside,pra,superscriptaddress]{revtex4}
\usepackage[utf8]{inputenc}

\usepackage[pdftex]{graphicx,color}
\usepackage{amsmath}
\usepackage{amssymb}
\usepackage{epsfig}
\usepackage{amsthm}
\usepackage{mathtools}
\usepackage{tikz}
\usepackage{wrapfig}
\usepackage{comment}
\usepackage{simpler-wick}
\usepackage{simplewick}
\usepackage{latexsym,revsymb}
\usepackage[unicode=true,pdfusetitle,
bookmarks=true,bookmarksnumbered=false,bookmarksopen=false,
breaklinks=false,pdfborder={0 0 0},backref=false,colorlinks=false]
{hyperref}

\DeclareMathOperator*{\argmax}{argmax}

\newcommand{\ket}[1]{| #1 \rangle}

\newcommand{\vertiii}[1]{{\left\vert\kern-0.25ex\left\vert\kern-0.25ex\left\vert #1 
    \right\vert\kern-0.25ex\right\vert\kern-0.25ex\right\vert}}

\makeatletter
\newtheorem*{rep@theorem}{\rep@title}
\newcommand{\newreptheorem}[2]{%
\newenvironment{rep#1}[1]{%
 \def\rep@title{#2 \ref{##1}}%
 \begin{rep@theorem}}%
 {\end{rep@theorem}}}
\makeatother

\newtheorem{theorem}{Theorem}
\newreptheorem{theorem}{Theorem}

\newreptheorem{corollary}{Corollary}

\newreptheorem{lemma}{Lemma}

\begin{document}
\title{Tensor-network codes}
\author{Terry Farrelly}
\email{farreltc@tcd.ie}
\affiliation{ARC Centre for Engineered Quantum Systems, School of Mathematics and Physics, The University of Queensland, St Lucia, QLD, 4072, Australia}
\author{Robert J. Harris}
\affiliation{ARC Centre for Engineered Quantum Systems, School of Mathematics and Physics, The University of Queensland, St Lucia, QLD, 4072, Australia}
\author{Nathan A. McMahon}
\affiliation{Center for Engineered Quantum Systems, Dept.\ of Physics and Astronomy, Macquarie University, 2109 New South Wales, Australia}
\affiliation{ARC Centre for Engineered Quantum Systems, School of Mathematics and Physics, The University of Queensland, St Lucia, QLD, 4072, Australia}
\author{Thomas M. Stace}
\affiliation{ARC Centre for Engineered Quantum Systems, School of Mathematics and Physics, The University of Queensland, St Lucia, QLD, 4072, Australia}
\begin{abstract}
	Inspired by holographic codes and tensor-network decoders, we introduce tensor-network stabilizer codes which come with a natural tensor-network decoder.  These codes can correspond to any geometry, but, as a special case, we generalize holographic codes beyond those constructed from perfect or block-perfect isometries, and we give an example that corresponds to neither.  Using the tensor-network decoder, we find a threshold of $18.8\%$ for this code under depolarizing noise.  We also show that for holographic codes the exact tensor-network decoder (with no bond-dimension truncation) is efficient with a complexity that is polynomial in the number of physical qubits, even for locally correlated noise.
	\end{abstract}

\maketitle

%\section{Introduction}
Tensor networks are a powerful tool in several branches of physics \cite{BC17}, so it is not surprising that they have also found use in quantum error correction.  One of the first examples of this was the use of matrix-product states to approximate the maximum likelihood decoder (which is the optimal decoder) for the surface code \cite{BSV14}.  This particular algorithm (with some modifications) has been applied to biased noise \cite{TDC19,TBF18} and correlated noise \cite{CF18} on the surface code.  Another approach using tensor networks to decode the surface code involved representing code states by projected entangled pair states and then finding the effective channel due to noise on the logical degrees of freedom \cite{DP17,DP18}.  This allowed one to go beyond Pauli noise and consider less-studied but important error models.  An early general method for decoding using tensor networks involved representing the encoding unitaries by a tensor network \cite{FP14,FP14a}.

Tensor networks also arose in holographic error correcting codes, which are toy models of the AdS/CFT correspondence \cite{FYH15,LS15,HNQ16,Evenbly17,JGP19,JGP19a,KC19,OS20}.  The idea is that the bulk degrees of freedom of a system in two-dimensional hyperbolic space (corresponding to logical qubits) are encoded in the boundary degrees of freedom (corresponding to the physical qubits).  This is analogous to the situation in AdS/CFT, where a strongly coupled gravity theory in the bulk is equivalent to a conformal field theory on the boundary \cite{Harlow16}.  These holographic error correcting codes were shown to reproduce much of the entanglement structure expected in AdS/CFT \cite{FYH15}, e.g., the Ryu-Takayangi formula \cite{RT06}.  

The holographic codes of \cite{FYH15} were constructed out of a type of isometry called perfect tensors, which is a strong constraint.  Afterwards, this class of tensors was replaced by the larger class of block perfect tensors (or equivalently perfect tangles) \cite{HMBS18,BO18}.  This led to the construction of CSS holographic codes \cite{HMBS18}, with one example constructed from the seven-qubit Steane code \cite{Steane96}, which was decoded with an integer optimization algorithm \cite{HC20} for Pauli noise (though the decoder is not efficient).

In this work, we introduce tensor-network stabilizer codes, which allow us to construct larger stabilizer codes out of smaller ones.  These codes naturally come with a tensor-network decoder (which is different but can be related to the tensor-network decoder of \cite{FP14}), which calculates the relevant probabilities for the maximum likelihood decoder.  Special cases of tensor-network codes include concatenated and holographic codes.  Here we use tensor-network codes to generalize holographic codes beyond perfect or block-perfect isometries.  We give an example corresponding to neither perfect nor block perfect isometries and which we decode using the tensor-network decoder to find a threshold of $18.8\%$ under depolarizing noise.  This compares quite well to the corresponding threshold for the surface code of 18.9\% \cite{BAO12}.  We also show that the exact tensor-network decoder is computationally efficient for holographic codes, with a polynomial run time in the number of physical qubits.

%\section{Stabilizer codes}
%\label{sec:Stabilizer}
\textit{Stabilizer codes}---In a stabilizer code \cite{Gottesman97,Gottesman09,NielsenChuang,Roffe19}, logical operators and stabilizers are elements of the $n$-qubit Pauli group $\mathcal{G}_n$, which consists of all operators like $z\sigma^{i_1}\otimes...\otimes\sigma^{i_n}$, where $z\in\{\pm 1,\pm i\}$.  Here $\sigma^{0}=\openone$, $\sigma^{1}=X$, $\sigma^{2}=Y$ and $\sigma^{3}=Z$, which are the single-qubit identity operator and three Pauli operators respectively.

We consider Pauli noise on the physical qubits, meaning that errors are elements of the Pauli group $\mathcal{G}_n$.  For the example holographic code, we will consider depolarizing noise on each of the physical qubits, where each physical qubit is subject to the quantum channel
\begin{equation}\label{eq:depnoise}
 D(\rho_1)=(1-p)\rho_1+\frac{p}{3}\sum_{i=1}^3\sigma^i\rho_1\sigma^i,
\end{equation}
where $0\leq p\leq 1$ is the probability of an error affecting the qubit.

The subspace corresponding to the logical information, the codespace, is fixed by an abelian group of operators $\mathcal{S}\subset\mathcal{G}_n$ called stabilizers.  The group is chosen so that, for any state $\ket{\psi}$ in the code space, $S\ket{\psi}=\ket{\psi}$ for every $S\in\mathcal{S}$.  The stabilizer group has $n-k$ independent generators $S_i$, so the logical subspace has dimension $2^k$ corresponding to $k$ logical qubits \cite{NielsenChuang}.

\begin{table}
\begin{center}
  \begin{tabular}{| c | c c c c c c c | }
    \hline
    Qubit & $0$ & 1 & 2 & 3 & 4 & 5 & 6 \\ \hline
    $S_1$ & $\openone$ & $Z$ & $\openone$ & $Z$ & $\openone$ & $\openone$ & $\openone$ \\ \hline
    $S_2$ & $\openone$ & $X$ & $Z$ & $Y$ & $Y$ & $X$ & $\openone$ \\ \hline
    $S_3$ & $\openone$ & $X$ & $X$ & $X$ & $X$ & $Z$ & $\openone$ \\ \hline
    $S_4$ & $\openone$ & $\openone$ & $Z$ & $Z$ & $X$ & $\openone$ & $X$ \\ \hline
    $S_5$ & $\openone$ & $X$ & $Y$ & $X$ & $Y$ & $\openone$ & $Z$ \\ \hline
    $X_1$ & $X$ & $X$ & $Z$ & $X$ & $Z$ & $\openone$ & $\openone$  \\ \hline
    $Z_1$ & $Z$ & $X$ & $Y$ & $Y$ & $X$ & $\openone$ & $\openone$  \\ 
    \hline
  \end{tabular}
\end{center}
	\caption{Stabilizer generators and logical operators for the $[[6,1,3]]$ code from \cite{SWO08}.  Their action on the physical qubits is shown in columns $1-6$, and their action on the logical codespace is shown in column $0$.  We can also think of these seven operators as stabilizing a state on all seven qubits $0-6$.}
\label{table:Steane}
\end{table}

Logical operators on the encoded qubits form a non-abelian group $\mathcal{L}\subset\mathcal{G}_n$.  This group is generated by the $k$ $X$-type and $k$ $Z$-type operators, which we may call $X_{\alpha}$ and $Z_{\alpha}$, where $\alpha\in\{1,..,k\}$.  These commute with the stabilizers and satisfy $X_{\alpha}Z_{\beta}=(-1)^{\delta_{\alpha \beta}}Z_{\alpha}X_{\beta}$.

The final important (abelian) group of operators are pure errors $\mathcal{E}\subset\mathcal{G}_n$.  This group is generated by $n-k$ operators $E_i$ which satisfy $E_iS_j=(-1)^{\delta_{ij}}S_jE_i$.  Note that the entire Pauli group is generated by all $E_i$, $S_i$, $X_{\alpha}$ and $Z_{\alpha}$.

To detect errors, we measure the stabilizer generators $S_i$ giving eigenvalues $s_i=\pm1$.  The syndrome $\vec{s}$ is the collection of these measurement outcomes.  We denote the pure error corresponding to syndrome $\vec{s}$ by $E(\vec{s}\,)$, meaning $E(\vec{s}\,)$ is the product of all $E_i$ such that $s_i=-1$.
$E(\vec{s}\,)$ is not the only error with syndrome $\vec{s}$.  Any error $E^{\prime}=LSE(\vec{s}\,)$, for any $L\in\mathcal{L}$ and any $S\in\mathcal{S}$, will have the same syndrome.  Deciding which operation to apply to correct whichever (if any) error has occurred is called decoding, and it is a difficult problem in general \cite{HG11,IP13}.

%\section{Tensor-network error correcting codes}
%\label{sec:TNcodes}
\begin{figure} [ht!]
\includegraphics[width=\columnwidth]{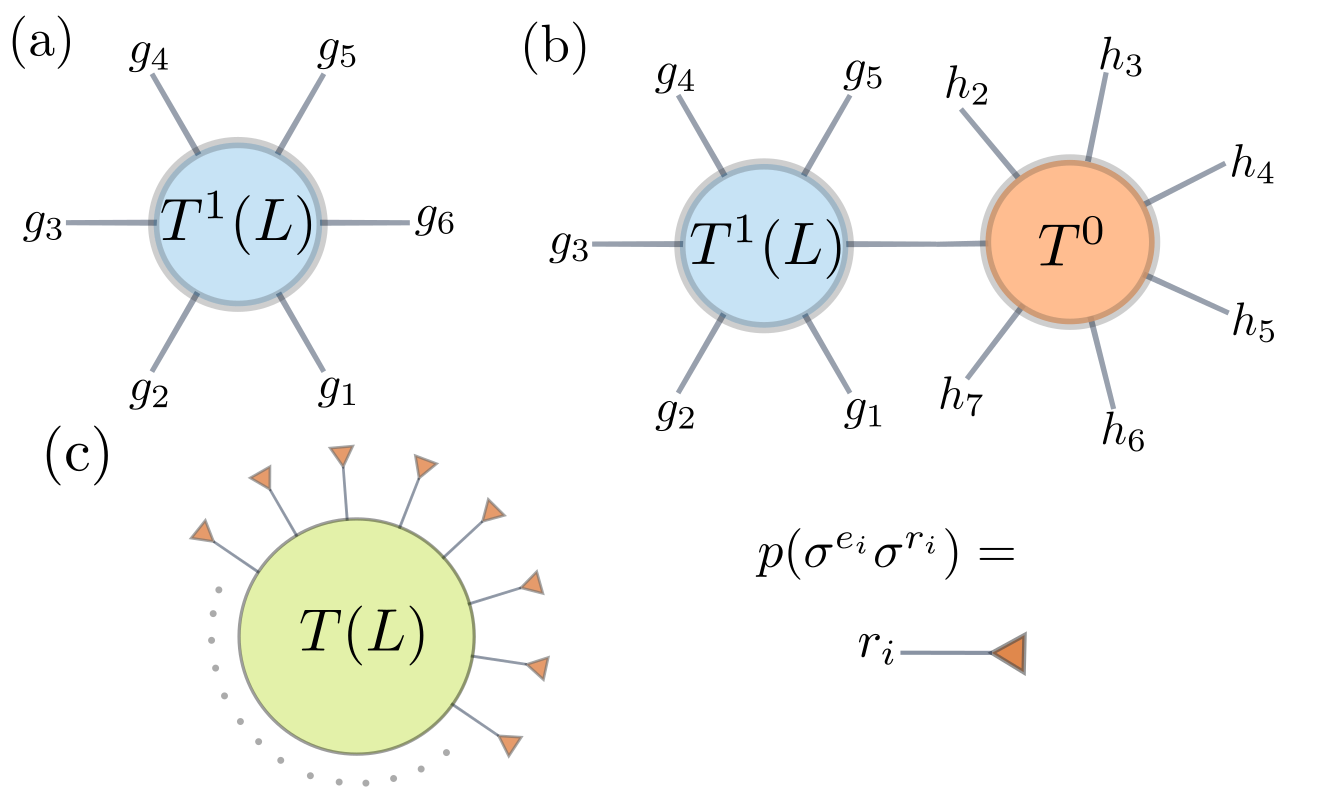}
	 \caption{Tensor-network codes:\ (a) the tensor $T^1(L)_{(g_1,...,g_6)}$ describing a six-qubit code, where $L$ denotes the logical operators.
	 (b) applying theorem \ref{th:first} to build a new stabilizer-code tensor by contracting the the sixth index of $T^1(L)_{(g_1,...,g_6)}$ with the first index of $T^0_{(h_1,...,h_7)}$ to get a $[[11,1,3]]$ code.
	 (c) given syndrome $\vec{s}$, the maximum-likelihood decoder finds the correction operator most likely to correct whichever error occurred.  A tensor-network decoder calculates a probability for each logical operator $L$ (the correction operator is determined by finding the maximimum of these probabilities) by contracting the code tensor with rank-one tensors $p(\sigma^{e_i}\sigma^{r_i})$ for each physical qubit $i$, with index $r_i$ ($e_i$ is fixed by the syndrome and so is not summed over).} 
\label{fig:TN_basics}
\end{figure}

\textit{Tensor-network error correcting codes}---Let us introduce tensors describing arbitrary stabilizer codes.  We represent operators by strings of integers, e.g., the stabilizer $XZYYX\openone=\sigma^{1}\sigma^{3}\sigma^{2}\sigma^{2}\sigma^{1}\sigma^{0}$ is represented by the string $(1,3,2,2,1,0)$. Then for each logical operator $L\in \mathcal{L}$, we define the rank-$n$ tensor (which is the indicator function for all operators in the class L)
\begin{equation}\label{eq:STdef}
	T(L)_{(g_1,...,g_n)}=\begin{cases}
                     1 \mathrm{\ \ \  if\ }\sigma^{g_1}\otimes...\otimes\sigma^{g_n}\in \mathcal{S}L\\
                     0 \mathrm{\ \ \  otherwise,}
                    \end{cases}
\end{equation}
where $g_j\in\{0,1,2,3\}$ and $\mathcal{S}L$ is the set of all operators of the form $SL$ with $S\in\mathcal{S}$.  In other words, $\mathcal{S}L$ is the coset of $\mathcal{S}$ with respect to the logical operator $L$.  For example, $T(\openone)_{(g_1,...,g_n)}$ is non zero only when $\sigma^{g_1}\otimes...\otimes\sigma^{g_n}$ is a stabilizer, so $T(\openone)$ describes the stabilizer group (except for the overall signs of the stabilizers, but once these are fixed for the generators, they are fully determined for the whole group).  Similarly, for a code with a single logical qubit, $T(X)_{(g_1,...,g_n)}$ describes the class of the logical $X$ operator.
 
These tensors are \emph{not} isometries, rather they describe the structure of the code in a simple way.  The tensors are agnostic about the encoding unitaries, but they can be related to encoding unitaries and the tensors of \cite{FP14} (see appendix \ref{app:FP}).

As an example, take the six-qubit code of \cite{SWO08}, which encodes one logical qubit into six physical qubits with stabilizer generators and logical operators summarised in table \ref{table:Steane}.  
The tensors for this code $T^1(L)_{(g_1,...,g_6)}$ have $32$ nonzero values for each possible $L\in\{\openone,X,Y,Z\}$, e.g., $T^1(X)_{(131300)}=1$.  We can define another code with no logical qubits (i.e., a stabilizer state) on seven physical qubits by taking the six-qubit code plus an extra qubit $0$ with all operators in table \ref{table:Steane} as stabilizers.  Let us denote this code tensor by $T^0_{(g_1,...,g_7)}$, so, e.g., $T^0_{(3122100)}=1$.
 
The benefit of describing codes by these code tensors is that we can combine several tensors together by contracting tensor legs to get new stabilizer codes, which come with a natural tensor-network decoder.
If the tensor network can be efficiently contracted, then the code can also be efficiently decoded.  We will see an example of this for holographic codes, where the decoder is efficient even without any approximations.
 The following theorem explains how to join code tensors to get new stabilizer codes (also see figure \ref{fig:TN_basics}).
\begin{theorem}\label{th:first}
	Consider two stabilizer-code tensors $T(L)_{(g_1,...,g_{n})}$ and $T^{\prime}(L^{\prime})_{(h_1,...,h_{n^{\prime}})}$ which have $n$ and $n^{\prime}$ physical qubits and $k$ and $k^{\prime}$ logical qubits respectively.  We can get new tensors describing a new stabilizer code by contracting indices (for simplicity, choose qubits $1$ to $l$ for both codes), i.e.,
 \begin{equation*}
 \begin{split}
	 & T_{\mathrm{new}}(L\otimes L^{\prime})_{(g_{l+1},...,g_{n},h_{l+1},...,h_{n^{\prime}})} = \\
	 & \sum_{j_1,...,j_l\in\{0,1,2,3\}}\!\!\!\!\!\!\!\!\!\!\! T(L)_{(j_1,...,j_{l},g_{l+1},...,g_{n})}T^{\prime}(L^{\prime})_{(j_1,...,j_{l},h_{l+1},...,h_{n^{\prime}})},
\end{split}
 \end{equation*}
	provided either one of these codes can distinguish any Pauli error on qubits $1$ to $l$.  $T_{\mathrm{new}}$ describes a stabilizer code with $n+n^{\prime} - 2l$ physical qubits and $k+k^{\prime}$ logical qubits.  (This is proved in \ref{app:th1_proof}.)
\end{theorem}

This tells us that we can build larger codes using small tensors as building blocks.  The tensor network can have any geometry and the component tensors can have any number of logical qubits.  It is easy to check if a tensor can be contracted onto another tensor to get a new stabilizer code:\ we just check if one of them can distinguish all different Pauli errors on the qubits corresponding to the contracted legs, which is equivalent to there existing an isometry from those legs (plus the logical qubits) to the rest (see appendix \ref{app:th1_proof}).  Theorem \ref{th:first} allows us to iteratively build up very large codes with a guarantee of consistency.  And, as we will see, it is exactly these tensors $T(L)$ that are contracted in the tensor-network decoder.  Also, it is straightforward to find stabilizer generators and pure errors (see appendix \ref{app:th1_proof}).

For example, we can contract index $g_6$ of $T^1(L)_{(g_1,...,g_{6})}$ and index $h_1$ of $T^0_{(h_1,...,h_{6})}$ as shown in figure \ref{fig:TN_basics}.  The six-qubit code can distinguish any single-qubit error on its sixth qubit (see table \ref{table:Steane}), which means that the resulting tensor
describes a stabilizer code with $11$ physical qubits and one logical qubit.

Another simple example are concatenated codes, e.g., for the six-qubit code:
\begin{equation*}
\begin{split}
	& T^{\mathrm{conc}}(L)_{(h_1,...,h_{36})}=\\
	& T^1(L)_{(g_1,...,g_6)}T^0_{(g_1,h_1,...,h_6)}T^0_{(g_2,h_7,...,h_{12})}\times\\
	& T^0_{(g_3,h_{13},...,h_{18})} T^0_{(g_4,h_{19},...,h_{24})}T^0_{(g_5,h_{25},...,h_{30})}T^0_{(g_6,h_{31},...,h_{36})},
 \end{split}
\end{equation*}
where repeated indices are contracted.  An interesting example of tensor-network codes will be holographic codes. 

Finally, these tensor-network codes should not be confused with quantum tensor product codes \cite{FLH17}, where one constructs codes by taking tensor products of the corresponding parity-check matrices.  Furthermore, they include, but are more general than, generalized concatenated codes \cite{GSS09,WZG13}.

%\section{Maximum likelihood decoding via tensor networks}
%\label{sec:Dec}
\textit{Maximum likelihood decoding via tensor networks}---The optimal decoder for quantum error correction is the maximum likelihood decoder, which finds the error correction operator that is most likely to return the system to the correct code state, given the syndrome (and an assumed error model).  

We know that any error corresponding to syndrome $\vec{s}$ has the form $E(\vec{s}\,)SL$ for some $L\in\mathcal{L}$ and $S\in\mathcal{S}$.  
Because $E(\vec{s}\,)SL$ has the same effect on the codespace for any $S$, we need to calculate
\begin{equation}\label{eq:1}
 \chi(L,\vec{s}\,)=\sum_{S\in\mathcal{S}} \mathrm{prob}(E(\vec{s}\,)SL)
\end{equation}
for each $L\in\mathcal{L}$, where $\mathrm{prob}(E(\vec{s}\,)SL)$ is the probability that the error $E(\vec{s}\,)SL$ occurs on the physical qubits.  Then the correction operator we should apply is $\overline{L}E(\vec{s}\,)$, where $\overline{L}=\argmax_L\chi(L,\vec{s}\,)$.

%\subsection{Tensor-network approach}
%\label{sec:TN_approach}
%\textit{Tensor-network approach}---
Calculating $\chi(L,\vec{s}\,)$ for a large number of physical qubits is difficult in general \cite{IP13}.  Luckily, in some cases, it is possible to write $\chi(L,\vec{s}\,)$ in terms of a tensor network that can be contracted efficiently.  
This idea was introduced for the surface code in \cite{BSV14} and for other codes via a circuit description in \cite{FP14}.
Writing an error as $E(\vec{s}\,)SL=\sigma^{a_1}\otimes...\otimes\sigma^{a_n}$ with $a_i\in\{0,1,2,3\}$, 
then, for the case of independent noise on each qubit, $\mathrm{prob}(\sigma^{a_1}\otimes...\otimes\sigma^{a_n})  = p_1(\sigma^{a_1})\times...\times p_n(\sigma^{a_n})$, where $p_i(\sigma^{a_i})$ is the probability that $\sigma^{a_i}$ will act on qubit $i$ due to the noise. For i.i.d.\ depolarizing noise, defined in equation (\ref{eq:depnoise}),  $p(\sigma^{a_i})=(1-p)\delta_{0a_i}+p/3(1-\delta_{0a_i})$.
If we write $E(\vec{s}\,)=\sigma^{e_1}\otimes...\otimes\sigma^{e_n}$, then we have
\begin{equation}
\begin{split}\label{eq:2}
	\chi(L,\vec{s}\,) & =\sum_{r_1,...,r_n\in\{0,1,2,3\}}T(L)_{(r_1...r_n)}\prod_{i=1}^{n} p_i(\sigma^{e_i}\sigma^{r_i}),
 \end{split}
\end{equation}
where $T(L)_{(r_1...r_n)}$ is the stabilizer-code tensor defined in equation (\ref{eq:STdef}).  This tensor network contraction is sketched in figure \ref{fig:TN_basics}.  We should think of $p(\sigma^{e_i}\sigma^{r_i})$ as a one-leg tensor associated to physical qubit $i$, with leg index $r_i$, as in figure \ref{fig:TN_basics}.  Note that $e_i$ is fixed because the pure error $E(\vec{s}\,)$ is fixed by the syndrome.  More generally, in the case of correlated noise, $\prod_{i=1}^{n} p_i(\cdot)$ will be replaced by $\mathrm{prob}(\cdot)$, though for the tensor network to be efficiently contractible, one would need some restrictions on $\mathrm{prob}(\cdot)$, such as a finite correlation length, e.g., factored noise \cite{CF18}.
\begin{figure*}[ht!]
\includegraphics[width=\textwidth]{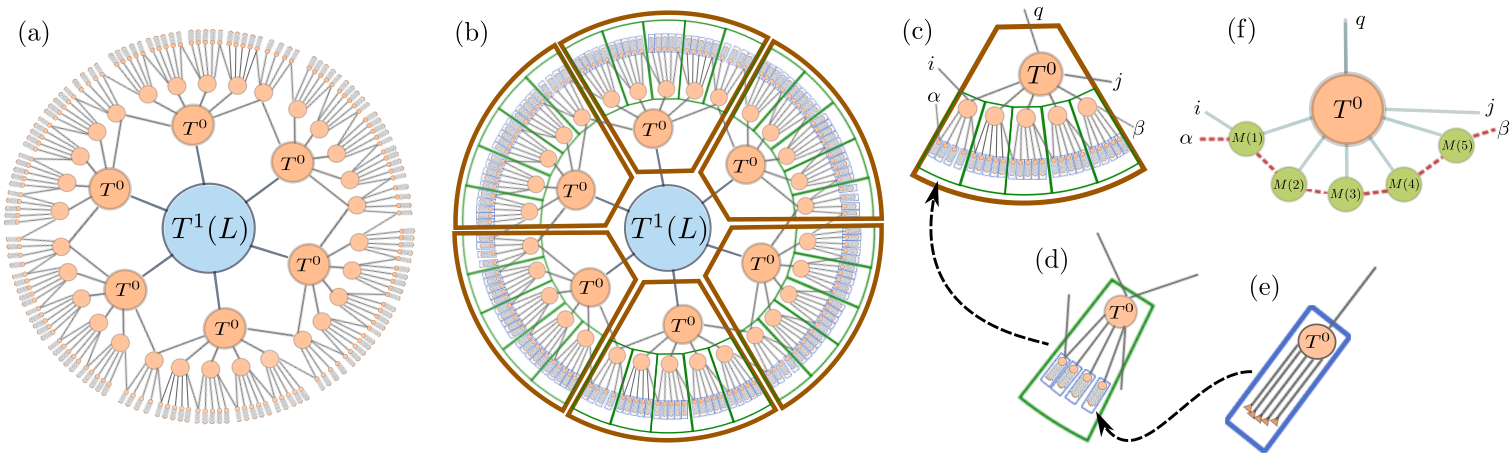}
	\caption{(a) the tensor network for a radius-four holographic code.  The central tensor $T^1(L)$ (corresponding to one logical qubit) has six legs whereas all other tensors $T^0$ with no logical qubits have seven legs.
	(b) the contraction order:\ starting from the outside, we contract inwards.  After the first contractions (e), we have tensors at radius four, which are enclosed inside the blue boxes.  After the next contraction (d), we have tensors at radius three, enclosed in the green boxes.  After another round of contractions (c), we have the tensor enclosed in the brown box.  Finally, these tensors are contracted with the central tensor, coloured blue in (b).  (f) the generic form of each tensor contraction, where $M(k)$ are the results of previous contractions, with bond legs marked by dashed lines.}
\label{fig:holo_TN}
\end{figure*}

For large codes the structure of $T(L)_{(r_1...r_n)}$ can be extremely complex and the contraction quickly becomes difficult with increasing $n$.  In \cite{BSV14} for the surface code the strategy was to decompose the tensor $T(L)$ (which described a code with only one logical qubit) into smaller tensors.  In \cite{CF18}, maximum likelihood decoding is recast as a calculation of partition functions which can be viewed as tensor network contractions, which is also applied to the surface code.  In \cite{FP14}, the encoding unitary circuit was represented by tensors, but the method can be related to that in equation (\ref{eq:2}) (see appendix \ref{app:FP}).  Our approach is to build tensor-network codes out of many smaller code tensors (which may have any number of logical qubits) with the goal that the contraction may be done efficiently.  We will now consider an example for holographic codes.

%\section{Holographic codes as tensor-network codes}
%\label{sec:holo_TN}
\textit{Holographic codes as tensor-network codes}---Using the tensor-network codes we can generalize previous incarnations of holographic codes, which relied on perfect \cite{FYH15} or block-perfect tensors \cite{HMBS18} or equivalently perfect tangles \cite{BO18}.  Let us consider an example using the six-qubit code as a building block, which is neither perfect nor block perfect (for any ordering of the indices).

This is best understood via figure \ref{fig:holo_TN}.  We start with a central six-qubit code tensor $T^1(L)$.  Then we contract each outgoing leg of this with the six tensors $T^{0}$ at radius one, each of which has one ingoing leg contracted with an outgoing leg of the central tensor.  We call this a radius-two code.  To get a radius-three code, we contract $T^0$ tensors with each outgoing leg of the radius-two tensors, but now some radius-three tensors have two legs contracted with two neighbouring radius-two tensors, as shown in figure \ref{fig:holo_TN}.  This pattern repeats until we reach the code radius $R$, at which point any external legs correspond to physical qubits.  Due to theorem \ref{th:first} we know that this is a valid stabilizer code, because $T^{0}$ can distinguish two-qubit errors on the qubits corresponding to the ingoing legs.  (We choose the leg ordering so that index $g_7$ of each tensor $T^{0}_{(g_1,...,g_7)}$ at radius $r+1$ is contracted with a tensor at radius $r$ except for those tensors at radius $r+1$ with two legs contracted with tensors at radius $r$ in which case we choose $g_6$ and $g_7$.)

%\subsection{Contracting the tensor network}
%\textit{Contracting the tensor network}---
To decode these holographic codes, we contract the tensor network to calculate $\chi(L,\vec{s}\,)$ for each $L\in\{\openone, X,Y,Z\}$.  As shown in figure \ref{fig:holo_TN}, we contract from the outside of the network in.  The bond dimension $D[r]$ of the tensors at radius $r$ grows like $D[r] = 4^{R-r}$ as we contract inwards, where $R$ is the code radius.

Even in the absence of any bond-dimension truncation scheme, the tensor-network contraction method here is efficient, meaning the number of operations is polynomial in the number of physical qubits $n$.  The total number of operations $N$ satisfies (see appendix \ref{app:complexity}) 
\begin{equation}\label{eq:complexity}
	N = bn^{\max(n_\mathrm{mat}/c,1)},
\end{equation}
where $c$ and $b$ are constants, and we have assumed the complexity of multiplying two $N\times N$ matrices is $O(N^{n_{\mathrm{mat}}})$, so $n_{\mathrm{mat}}\simeq 2.37$ in the best case \cite{Gall14}.  For holographic codes with tensors having more than seven legs (as is the case for the example we consider), $c\geq 1$.  Otherwise, $c$ is constant but between zero and one.
This should be contrasted with the method of \cite{BSV14} for the surface code, which is also polynomial in the number of physical qubits, but which uses bond-dimension truncation to approximate the maximum likelihood decoder.
\begin{figure}[ht!]
\includegraphics[width=\columnwidth]{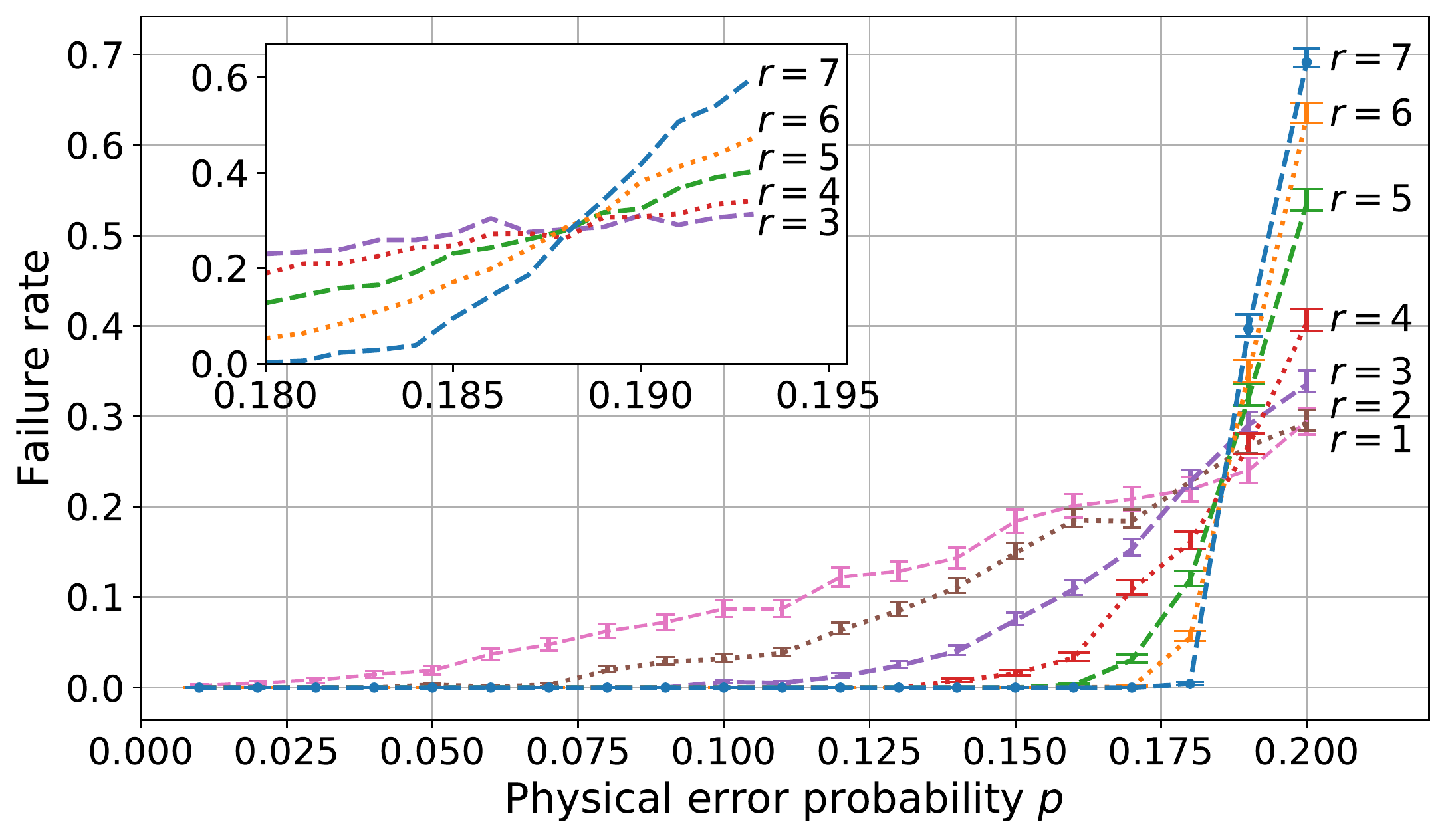}
	\caption{Failure rate for the zero-rate six-qubit holographic code as a function of the single-qubit error probability (depolarizing noise parameter $p$) for different code radii.  Inset:\ expanded view of the failure rate close to the threshold, estimated to be $18.8\%$ (see appendix \ref{app:threshold}).}
\label{fig:thresh}
\end{figure}

We applied this contraction algorithm to our example holographic code to find out how well it performs under depolarizing noise. Monte Carlo simulation results are shown in figure \ref{fig:thresh}.  We see a threshold for the code under depolarizing noise at $18.8\%$ (see appendix \ref{app:threshold}), which compares well with the threshold for the surface code of $18.9\%$ \cite{BAO12}.

%\section{Conclusions}
\textit{Conclusions}---We described stabilizer codes by tensors and showed that we can combine codes by contracting tensor legs to get larger stabilizer codes.  These tensor-network stabilizer codes come with a natural tensor-network decoder.  We applied this to holographic codes generalizing previous constructions and found a new code with a threshold of $18.8\%$ under depolarizing noise.  We also found that the computational complexity of decoding holographic codes via the maximum likelihood decoder is polynomial in the number of physical qubits even in the exact case, with no bond-dimension truncation.  Furthermore, in practice even for a single contraction of the tensor network, many operations can be done in parallel, which would greatly speed up decoding.

%\acknowledgments
\textit{Acknowledgments}---The authors would like to thank Aidan Strathearn for useful discussions.
This work was supported by the Australian Research Council Centres of Excellence for Engineered Quantum Systems (EQUS, CE170100009) and the Asian Office of Aerospace Research and Development (AOARD) grant FA2386-18-14027.
Numerical simulations were performed on The University of Queensland's School of Mathematics and Physics Core Computing Facility ``getafix'' (with thanks to Dr. L. Elliott and I. Mortimer for computing support).

\bibliographystyle{unsrt}
%\bibliography{../../References}

\appendix
\section{Relation of code tensors to encoding unitaries}
\label{app:FP}
The code tensors $T(L)$, defined in equation (\ref{eq:STdef}), can be related to any encoding unitary as follows.  Consider a unitary $U$ that takes products of Pauli operators to products of Pauli operators (a Clifford unitary).  This can be represented as a simple tensor $u_{i_1,...,i_n}^{j_1,...,j_n}$, defined via
\begin{equation}
 U^{\dagger}(\sigma^{j_1}\otimes...\otimes\sigma^{j_n})U = \sum_{i_1,...,i_n\in\{0,1,2,3\}}u_{i_1,...,i_n}^{j_1,...,j_n}\sigma^{i_1}\otimes...\otimes\sigma^{i_n},
\end{equation}
with the property that, for each fixed set of indices $j_1,...,j_n$, the tensor $u_{i_1,...,i_n}^{j_1,...,j_n}\in\{1,-1,i,-i\}$ for exactly one set of indices $i_1,...,i_n$, and $u_{i_1,...,i_n}^{j_1,...,j_n}=0$ for all the rest.  Any stabilizer code state can be prepared by such a Clifford unitary.  We may take the first $k$ input qubits to correspond to the logical input qubits, so the image under $U$ of any operator of the form $\sigma^{j_1}\otimes...\otimes \sigma^{j_k}\otimes \sigma^0\otimes...\otimes \sigma^0$ is a logical operator on the code.

Similarly, we may take the stabilizer generators to be the image under $U$ of all operators $\sigma^3_l$, where $l\in\{k+1,...,n\}$, and $\sigma^3_l$ is a Pauli $Z$ operator on qubit $l$ (and acts as the identity on all other qubits).  Thus, we have that the code tensor $T(L)$ defined in equation (\ref{eq:STdef}) representing the logical coset $\mathcal{S}L$ for logical operator $L=\sigma^{j_1}\otimes...\otimes \sigma^{j_k}$ is given by
\begin{equation}
 T(\sigma^{j_1}\otimes...\otimes \sigma^{j_k})  =\sum_{j_{k+1},...,j_n\in\{0,3\}}|u_{i_1,...,i_n}^{j_1,...,j_n}|.
\end{equation}

The tensor-network decoder of \cite{FP14} calculates $\chi(L,\vec{s}\,)$ via (for independent noise on each physical qubit, with probabilities $p(\sigma^{i_{\alpha}})$)
\begin{equation}
 \chi(L,\vec{s}\,) =\sum_{\substack{
                             i_1,...,i_n\in\{0,1,2,3\}\\
                             j_{k+1},...,j_n\in\{0,1,2,3\}
                            }}|u_{i_1,...,i_n}^{j_1,...,j_n}|\prod_{\alpha=1}^{n} p(\sigma^{i_{\alpha}})\prod_{\beta=k+1}^{n} s_{j_{\beta}},
\end{equation}
where $L=\sigma^{j_1}\otimes...\otimes \sigma^{j_k}$, and we have the syndrome tensors $s_{j_{\beta}}=(1,0,0,1)$ if the syndrome corresponding to stabilizer $S_{\beta}$ is $+1$ and $s_{j_{\beta}}=(0,1,1,0)$ if the syndrome corresponding to $S_{\beta}$ is $-1$.

This is useful if $u_{i_1,...,i_n}^{j_1,...,j_n}$ can be decomposed into a product of smaller encoding unitaries and the resulting tensor-network contraction can be done efficiently.  In \cite{FP14,FP14a} examples of codes were considered where this was possible.

This approach is focussed on sequentially applying encoding unitaries (typically CNOTs), whereas in our approach we allow contraction of output legs from code tensors:\ e.g., $\sum_j T^1(L_1)_{g_1,...,g_5,j}T^1(L_2)_{j,h_2,...,h_6}$ describes a stabilizer code with two logical qubits and ten physical qubits.  This can in principle also be reinterpreted as a sequence of two encoding unitaries, but this will become very cumbersome in contrast to the simple formula in our case because we can easily contract two output legs of, e.g., the same code tensor, whereas with unitaries, one needs to chain different unitaries.  Also, our approach is agnostic about the precise form of the encoding circuit.  Furthermore, our decoder also allows locally correlated noise, which is no obstacle to efficiency in the case of holographic codes.  

\section{Proof of theorem 1}
\label{app:th1_proof}
Suppose we have the two stabilizer-code tensors $T(L)_{g_1,...,g_{n}}$ and $T^{\prime}(L^{\prime})_{h_1,...,h_{n^{\prime}}}$ which have $n$ and $n^{\prime}$ physical qubits and $k$ and $k^{\prime}$ logical qubits respectively.  Our goal is to show that the tensors
 \begin{equation}\label{eq:app1}
 \begin{split}
& T_{\mathrm{new}}(L\otimes L^{\prime})_{g_{l+1},...,g_{n},h_{l+1},...,h_{n^{\prime}}} = \\
& \sum_{j_1,...,j_l\in\{0,1,2,3\}}\!\!\!\!\!\!\!\!\!\!\! T(L)_{j_1,...,j_{l},g_{l+1},...,g_{n}}T^{\prime}(L^{\prime})_{j_1,...,j_{l},h_{l+1},...,h_{n^{\prime}}}
\end{split}
 \end{equation}
 describe a new stabilizer code provided that at least one of the codes $T(L)$ or $T^{\prime}(L^{\prime})$ can distinguish any Pauli error on the set of qubits corresponding to the contracted legs.  
 
 For simplicity, first we will show that $T_{\mathrm{new}}(L\otimes L^{\prime})$ describes a stabilizer code in the case where $l=1$ (i.e., we are only contracting one index), and then at the end we explain how to generalize this to arbitrary $l$.  Also, without loss of generality, we will assume that it is tensor $T(L)$ that has the property that it can distinguish arbitrary Pauli errors on the qubits corresponding to the contracted legs.
 
For $T_{\mathrm{new}}(L\otimes L^{\prime})$ to describe a stabilizer code, we need it to have the following three properties:\
\begin{enumerate}
 \item[(i)] The elements of $T_{\mathrm{new}}(L\otimes L^{\prime})$ must be either zero or one as in equation (\ref{eq:STdef}).
 \item[(ii)] $T_{\mathrm{new}}(\openone\otimes \openone^{\prime})$ must describe a stabilizer group of $n+n^{\prime}-2$ physical qubits with $k+k^{\prime}$ logical qubits.
 \item[(iii)] $T_{\mathrm{new}}(L\otimes L^{\prime})$ must describe the logical coset of $L\otimes L^{\prime}$ with the right commutation relations between all representatives of logical operators.  E.g., all stabilizers should commute, meaning all operators described by $T_{\mathrm{new}}(\openone\otimes \openone^{\prime})$ should commute.
\end{enumerate}

%\begin{center}
\textbf{Proof of (i):}
%\end{center}
For any set of indices the tensor $T_{\mathrm{new}}(L\otimes L^{\prime})$ takes positive integer values, which can be seen from the sum in equation (\ref{eq:app1}) and noting that the entries of $T(L)$ and $T^{\prime}(L^{\prime})$ are either zero or one.  Then we can show that the entries of $T_{\mathrm{new}}(L\otimes L^{\prime})$ can only be zero or one by contradiction.  Suppose an entry of $T_{\mathrm{new}}(L\otimes L^{\prime})$ is greater than one.  This implies that, for some value of the indices $g_{2},...,g_{n}$ and some $L$, the tensor $T(L)_{k,g_{2},...,g_{n}}=1$ for more than one value of the index $k$.  Then both $\sigma^k\otimes \sigma^{g_{2}}\otimes...\otimes\sigma^{g_{n}}$ and $\sigma^{k^{\prime}}\otimes \sigma^{g_{2}}\otimes...\otimes\sigma^{g_{n}}$ with $k\neq k^{\prime}$ must describe stabilizers or logical operators.  By taking the products of these two operators we find that $\sigma^{k^{\prime\prime}}\otimes\openone\otimes...\otimes\openone$ is also a stabilizer or logical operator (for some $k^{\prime\prime}\neq 0$).  But any code with this property cannot detect the single-qubit error $\sigma^{k^{\prime\prime}}$ on qubit $1$, since every stabilizer must commute with $\sigma^{k^{\prime\prime}}\otimes\openone\otimes...\otimes\openone$, so we have a contradiction.

%\begin{center}
\textbf{Proof of (ii):}
%\end{center}
To show that property (ii) is satisfied, it is useful to think in terms of the stabilizers of $T(L)$ and $T^{\prime}(L^{\prime})$ before contracting the tensor indices.  We need to find which stabilizers survive the contraction:\ exactly those that match on qubits $1$ of $T(L)$ and $1^{\prime}$ of $T^{\prime}(L^{\prime})$.

Because $T(L)$ can distinguish an arbitrary error on qubit $1$, we may choose its stabilizer generators to have a useful form.  We choose exactly one stabilizer generator $S_1$ to have a Pauli $X$ on qubit $1$ and exactly one stabilizer generator $S_2$ to have a Pauli $Z$ on qubit $1$ with the rest having identities on qubit $1$.  ($S_1$ and $S_2$ are analogous to the pushing vectors of \cite{FYH15}.)  For any stabilizer generator in $T^{\prime}(L^{\prime})$, $S_j^{\prime}$, which acts like $\sigma^k$ on its qubit $1^{\prime}$, we can choose $\mu_1,\mu_2\in\{0,1\}$ such that $P_j=(S_1)^{\mu_1}(S_2)^{\mu_2}$ acts like $\sigma^k$ on qubit $1$ of $T(L)$.  The only stabilizer generators (of the tensor product of both codes $T(L)$ and $T^{\prime}(L^{\prime})$) that survive the index contraction are then $P_j\otimes S_j^{\prime}$ for all $j$ and $S_i\otimes\openone^{\prime}$ for $i\in\{3,..,n-k\}$.  This means that there are two fewer stabilizer generators after contraction.  Before contraction, the stabilizer group of the two codes had $n-k+n^{\prime}-k^{\prime}$ generators, and after contraction the stabilizer group of the new code has $n-k+n^{\prime}-k^{\prime}-2$ generators.  In the new code, there are $n+n^{\prime}-2$ physical qubits, so the number of logical qubits must be $k+k^{\prime}$.

%\begin{center}
\textbf{Proof of (iii):}
%\end{center}
Consider any representatives for the two operators $L_{1}\otimes L_{1}^{\prime}$ and $L_{2}\otimes L_{2}^{\prime}$ on the new code, which we can write as $\sigma^{\underline{a}}\otimes \sigma^{\underline{b}}$ and $\sigma^{\underline{c}}\otimes \sigma^{\underline{d}}$, where $\sigma^{\underline{a}}$ is shorthand for $\sigma^{a_{2}}\otimes...\otimes \sigma^{a_{n}}$, which acts on the first $n-1$ qubits and $\sigma^{\underline{b}}$ is shorthand for $\sigma^{b_{n}}\otimes...\otimes \sigma^{b_{n^{\prime}+n-1}}$, which acts on qubits $n$ to $n^{\prime}+n-2$.  These obey the algebraic relation
\begin{equation}
\begin{split}
& (L_{1}\otimes L_{1}^{\prime})(L_{2}\otimes L_{2}^{\prime}) = (\sigma^{\underline{a}}\otimes \sigma^{\underline{b}})(\sigma^{\underline{c}}\otimes \sigma^{\underline{d}})\\
&=z(\sigma^{\underline{c}}\otimes \sigma^{\underline{d}})(\sigma^{\underline{a}}\otimes \sigma^{\underline{b}})=z(L_{2}\otimes L_{2}^{\prime})(L_{1}\otimes L_{1}^{\prime})
\end{split}
\end{equation}
for some $z\in\{1,-1\}$.  If we look at the original codes before contraction, the corresponding operators are $(\sigma^l\otimes\sigma^{\underline{a}})\otimes (\sigma^l \otimes \sigma^{\underline{b}})$ representing $L_{1}\otimes L_{1}^{\prime}$ and $(\sigma^{k}\otimes \sigma^{\underline{c}})\otimes (\sigma^{k}\otimes\sigma^{\underline{d}})$ representing $L_{2}\otimes L_{2}^{\prime}$ for some $l,k\in\{0,1,2,3\}$.  These obey the same algebraic relation
\begin{equation}
\begin{split}
 & \left[(\sigma^l \otimes \sigma^{\underline{a}})\otimes (\sigma^l \otimes\sigma^{\underline{b}})\right]\left[(\sigma^{k} \otimes \sigma^{\underline{c}})\otimes (\sigma^{k} \otimes \sigma^{\underline{d}})\right]\\
 & =z\left[(\sigma^{k}\otimes \sigma^{\underline{c}})\otimes (\sigma^{k}\otimes \sigma^{\underline{d}})\right]\left[(\sigma^l \otimes \sigma^{\underline{a}})\otimes (\sigma^l \otimes \sigma^{\underline{b}})\right],
 \end{split}
\end{equation}
because $\sigma^l$ and $\sigma^{k}$ both appear twice, so whether they commute or anticommute does not matter.  In particular, this verifies that the stabilizer group, described by $T_{\mathrm{new}}(\openone\otimes\openone^{\prime})$, is indeed abelian.  It also implies that any representation of the logical group of operators has the correct algebraic structure, e.g., if each code $T(L)$ and $T^{\prime}(L^{\prime})$ had a single logical qubit, then, for example, $X\otimes Z^{\prime}$ would anticommute with $Y\otimes Z^{\prime}$.

%\begin{center}
\textbf{Finding representatives of operators:}
%\end{center}
In proving property (ii), we chose a useful form for the stabilizer generators of $T(L)$.  A consequence is that, by possibly taking products with these stabilizers, we can always find a representative of any logical operator $L$ that acts like the identity on qubit $1$ of code $T(L)$.  Suppose this had the form $\sigma^0\otimes A$, where $A$ is an operator on qubits $2$ to $n$ of code $T(L)$.  Then on the new code $T_{\mathrm{new}}(L\otimes L^{\prime})$, one representative of $L\otimes \openone$ will be $A\otimes \sigma^0\otimes...\otimes\sigma^0$, where the identities $\sigma^0$ act on qubits $n$ to $n+n^{\prime}-1$ of the new code.

For logical operators $L^{\prime}$ from the code $T(L^{\prime})$, we can first choose any representative, which we can write as $\sigma^k\otimes B$, where $B$ is an operator on qubits $2$ to $n^{\prime}$ of the code $T(L^{\prime})$.  We can always find $\mu_1,\mu_2\in\{0,1\}$ such that the product of stabilizers $Q=S_1^{\mu_1}S_2^{\mu_2}$ has a $\sigma^k$ on qubit $1$ of the code $T(L)$, which means that the operator $Q\otimes L^{\prime}$ survives the contraction, which gives us a representative of $\openone\otimes L^{\prime}$ on the new code.

A similar argument applies to pure errors, and the end result is that it is easy to find stabilizer generators, representatives of logical operators and pure errors for the new code, given the corresponding operators of the original codes.

%\begin{center}
\textbf{Generalization to many qubits:}
%\end{center}
If we are contracting more than one index, all the same arguments apply except that we need to choose the stabilizer generators to have the following form on the first $l$ qubits of the code $T(L)$.  We choose the stabilizer generators such that
 \begin{equation}\label{eq:stab-form}
  \begin{split}
   S_1 & =\sigma^1\otimes\openone...\openone\otimes A_1\\
   S_2 & =\sigma^3\otimes\openone...\openone\otimes A_2\\
   S_3 & =\openone\otimes\sigma^1\otimes\openone...\openone\otimes A_3\\
   S_4 & =\openone\otimes\sigma^3\otimes\openone...\openone\otimes A_4\\
   & \vdots\\
   S_{2l-1} & =\openone...\openone\otimes\sigma^1\otimes A_{2l-1}\\
   S_{2l} & =\openone...\openone\otimes\sigma^3\otimes A_{2l},
  \end{split}
 \end{equation}
 where $A_i$ are tensor products of Paulis on qubits $l+1$ to $n$.  This means that, for any combination of Pauli operators on the first $l$ qubits, there is a unique product of the first $2l$ stabilizers that has that form on those qubits.
 
The stabilizer generators can always be chosen to have this form if the code $T(L)$ can distinguish all Pauli errors on qubits $1$ to $l$.  To see this, consider the errors
 \begin{equation}
  \begin{split}
   E_1 & =\sigma^3\otimes\openone...\\
   E_2 & =\sigma^1\otimes\openone...\\
   & \vdots\\
   E_{2l-1} & =\openone...\openone\otimes\sigma^3\otimes \openone...\\
   E_{2l} & =\openone...\openone\otimes\sigma^1\otimes \openone....
  \end{split}
 \end{equation}
Given any set of stabilizer generators $S_i$, at least one will anticommute with $E_1$ (otherwise the code could not detect that error).  Relabel indices of the stabilizer generators such that $S_1$ anticommutes with $E_1$, and for any other generator $S_i$ that also anticommutes with $E_1$ make the replacement $S_i\rightarrow S_iS_1$, so that the only stabilizer that anticommutes with $E_1$ is $S_1$.  Repeat this procedure for $E_2$, so that the only stabilizer anticommuting with $E_2$ is $S_2$.  (In general, it is possible that the only stabilizer anticommuting with $E_2$ is also $S_1$, but we have assumed that the code can distinguish each different error, so this possibility cannot occur.)  We repeat this for each $E_i$ up to $i=2l$.  Finally, because, e.g., $S_1$ commutes with all $E_i$ except $E_1$, it must have the form $\sigma^1\otimes\openone...\openone\otimes A_1$, where $A_1$ is some tensor product of Paulis on qubits $l+1$ to $n$, and similarly for the other stabilizers.

\textbf{Isometries:} Now let us relate the condition that a code can distinguish all Pauli errors on some qubits to the existence of an isometry from those qubits (plus the logical qubits) to the rest.

Suppose the encoding isometry for a stabilizer code is given by
\begin{equation}
 V=\sum_{i_1,...,i_{n+k}\in\{0,1\}}J_{i_1,...,i_{n+k}}|i_{1},...,i_{n}\rangle\langle i_{n+1},...i_{n+k}|.
\end{equation}
Then, via the Choi–Jamiołkowski isomorphism, there is a corresponding stabilizer state given by
\begin{equation}
 |J\rangle = \sum_{i_1,...,i_{n+k}\in\{0,1\}}J_{i_1,...,i_{n+k}}|i_1,...,i_{n+k}\rangle.
\end{equation}
This state is stabilized by all stabilizers of the code, which act as $S_{1,...,n}\otimes\openone_{n+1,...,n+k}$, where $S_{1,...,n}$ is a stabilizer acting on the first $n$ qubits and $\openone_{n+1,...,n+k}$ is the identity on the last $k$ qubits.  Furthermore, the state $|J\rangle$ is also stabilized by operators formed from the logical operators of the original code.  These act as $L_{1,...,n}\otimes L^{\mathrm{log}}_{n+1,...,n+k}$, where $L^{\mathrm{log}}_{n+1,...,n+k}$ is a Pauli product on the last $k$ qubits, and $L_{1,...,n}$ is a representative of this Pauli product on the first $n$ qubits.

Now suppose that the code can distinguish all Pauli errors on qubits $n-j+1$ to $n$.  Via equation (\ref{eq:stab-form}), by taking products of stabilizer generators, we can find stabilizers realizing any product of Paulis on qubits $n-j+1$ to $n$.  Also, because the logical operators are now stabilizers for the state $|J\rangle$, there is also a stabilizer for every product of Paulis on qubits $n+1$ to $n+k$.

So consider a stabilizer of $|J\rangle$, given by $S=A\otimes B$, where $A$ is some product of Paulis on qubits $1$ to $n+j$, and $B$ is a product of Paulis on qubits $n-j+1$ to $n+k$.  We know that
\begin{equation}\label{eq:ABV}
 S|J\rangle=(A\otimes B) |J\rangle= |J\rangle.
\end{equation}

Now consider the new linear operator
\begin{equation}
 U = \sum_{i_1,...,i_{n+k}\in\{0,1\}}J_{i_1,...,i_{n+k}}|i_1,...,i_{n-j}\rangle\langle i_{n-j+1},...i_{n+k}|.
\end{equation}
 It follows from equation (\ref{eq:ABV}) that 
 \begin{equation}
	 AUB^{T}= U,
\end{equation}
which we used that the maximally entangled state in the Choi-Jamiołkowski isomorphism satisfies $\sum_{\alpha=1}^d\langle \alpha \alpha|M\otimes \openone = \sum_{\alpha=1}^d\langle \alpha \alpha|\openone\otimes M^{T}$ for any operator $M$.
Then it follows that
  \begin{equation}
	  B^{T}U^{\dagger}UB^{T}= (U^{\dagger}A)AU=U^{\dagger}U
\end{equation}
 since $A^{2}=\openone$, and $A$ and $B^T$ are self-adjoint.  But $B$ can be any product of Pauli operators on qubits $n-j+1$ to $n+k$, which means that $U^{\dagger}U\propto \openone_{n-j+1,...,n+k}$, so $U$ is proportional to an isometry.

\section{Proof of equation (\ref{eq:complexity})}
\label{app:complexity}
To prove equation (\ref{eq:complexity}), which bounds the complexity of the contraction algorithm, it helps to look at figure \ref{fig:holo_TN}.  Recall that the bond dimension grows with the radius we have contracted to $r$, i.e.,
\begin{equation}\label{eq:bond_dim}
 D[r] = 4^{R-r},
\end{equation}
for all $r\leq R$, where $R$ is the code radius.  We also have $D[R+1]=1$ as the tensors at radius $R+1$ are the single-leg tensors describing the probabilities of Pauli errors on the physical qubits.  Using the contraction scheme described in figure \ref{fig:holo_TN}, all contractions have the same form (except for the final contraction with the central tensor which is slightly different).

Here we will assume that each tensor in the holographic code has $m$ legs and we will denote them by $T^0$, but one could also consider codes with different tensors at each site (and which may have logical qubits too), in which case $m$ can be taken to be the maximum number of legs of any tensor in the code to get a similar bound.
\begin{figure} [ht!]
\includegraphics[width=6.7cm]{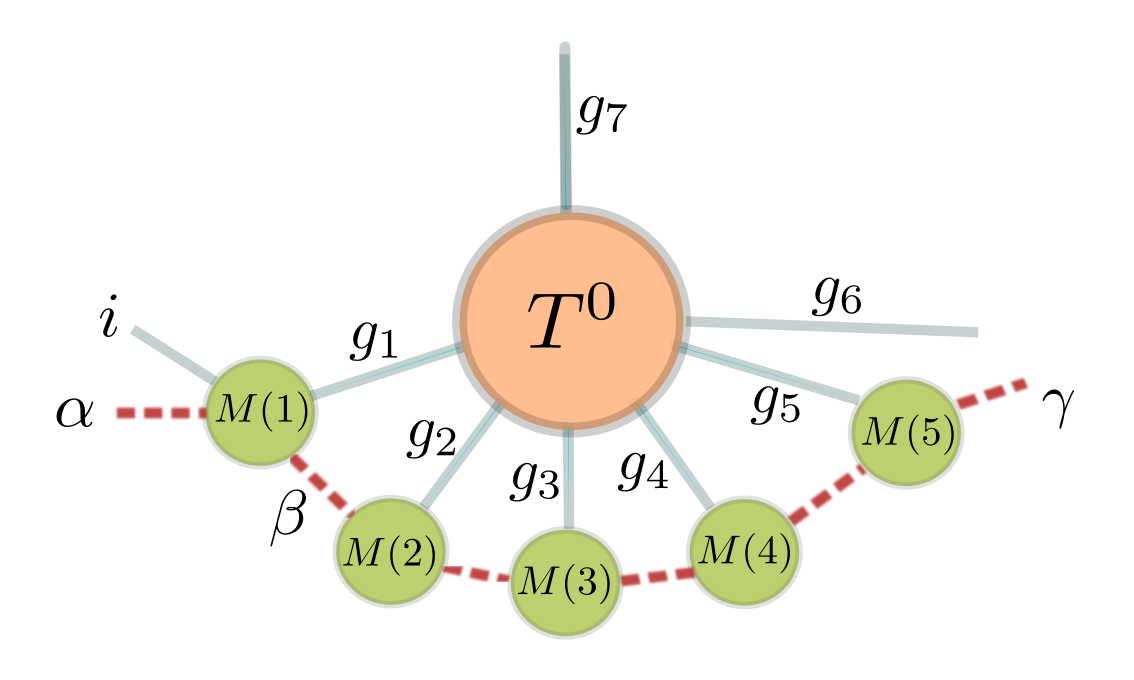}
	 \caption{Contracting the tensor network for holographic codes involves repeated application of contractions of the form illustrated above.  A tensor at radius $r$ is contracted with tensors at radius $r+1$ that may be the results from previous contractions.  In this example, the tensor at radius $r$ is denoted $T^0$ which has seven legs, but the ideas apply more generally.} 
\label{fig:order}
\end{figure}

Consider the tensor at radius $r$ (labelled with $T^0$ in figure \ref{fig:order}).  For each string of indices $g_1,...,g_m$ for which $T^0_{g_1,...,g_m}\neq 0$, we contract the other tensors (labelled by $M(k)$ with $k\in\{1,...,m-2\}$ in figure  \ref{fig:order}).  This entails multiplying a constant number of matrices.  As $g_1$ is now fixed, the first tensor $M(1)^{i\,g_1}_{\alpha\beta}$ consists of four $D[r+1]\times D[r+1]$ matrices for each value of index $i$ (in practice, it is easier to view it as a $D[r]\times D[r+1]$ matrix, though the bound we would get would be a bit more convoluted).  The other tensors $M(k)$ are $D[r+1]\times D[r+1]$ matrices.  So, since there are $m-2$ tensors $M(k)$, the complexity of contracting them is $4(m-3)\times c_{\mathrm{mat}}D[r+1]^{n_{\mathrm{mat}}}$, where we are assuming that multiplying two $N\times N$ matrices takes at most $c_{\mathrm{mat}}N^{n_{\mathrm{mat}}}$ operations.  The fastest known algorithm  has $n_{\mathrm{mat}}\simeq 2.37$ \cite{Gall14}, but typically different algorithms are used in practice with a slightly higher value of $n_{\mathrm{mat}}$.  We have to do this contraction for each nonzero value of $T^0_{g_1,...,g_m}$, which is a total of (at most) $2^m$ different contractions.  (This is because each code tensor has at most $2^m$ non-zero entries.)  Thus, we get a total of $2^{m+2}(m-3) c_{\mathrm{mat}} D[r+1]^{n_{\mathrm{mat}}}$ operations to do the contraction in figure \ref{fig:order}.

However, some of the tensors at radius $r$ will actually have \emph{two} legs joined to tensors at radius $r-1$ as opposed to one, which means there are only $m-3$ tensors $M(k)$ to contract, so these cases require fewer operations, i.e., $2^{m+2}(m-4)c_{\mathrm{mat}} D[r+1]^{n_{\mathrm{mat}}}$, but for the sake of simplicity, for each contraction we use the larger value.

For the final contraction with the central tensor, for each string of indices $g_1,...,g_m$ for which $T(L)_{g_1,...,g_m}\neq 0$, we take the trace of the product of tensors at radius two.  So we multiply $m$ matrices and take the trace, which takes at most $(m-1)c_{\mathrm{mat}} D[2]^{n_{\mathrm{mat}}}+D[2]$ operations.  Since we do this at most $2^m$ times, we get a total of $2^m((m-1)c_{\mathrm{mat}} D[2]^{n_{\mathrm{mat}}}+D[2])$ operations.  However, this is less than the general formula $2^{m+2}(m-3) c_{\mathrm{mat}} D[r+1]^{n_{\mathrm{mat}}}$ with $r=1$, so we can use the general expression as an upper bound.

Let us denote the number of tensors at radius $r$ by $n_{\mathrm{tens}}(r)$.
Then the total number of operations is bounded above by
\begin{equation}
N\leq K\sum_{r=R}^{1}n_{\mathrm{tens}}(r)D[r+1]^{n_\mathrm{mat}},
\end{equation}
where we defined $K = 2^{m+2}(m-3)c_{\mathrm{mat}}$.

Next we use that the number of tensors at each radius increases exponentially.  E.g., for the holographic code we are considering, each tensor at radius $r$ has at least five outgoing legs (to radius $r+1$), so there are at least four unique tensors at radius $r+1$ for each tensor at radius $r$.  (See figure \ref{fig:holo_TN}.)  So we choose the largest $\gamma>1$ (which is at least four for our example) such that
\begin{equation}
 n_{\mathrm{tens}}(r+1)\geq \gamma\, n_{\mathrm{tens}}(r)
\end{equation}
for all $r$.  This implies that
\begin{equation}
 n\geq \gamma n_{\mathrm{tens}}(R)\geq \gamma^{R-r+1} n_{\mathrm{tens}}(r)
\end{equation}
for all $r$.  Then with $c=\log_4(\gamma)$, we get
\begin{equation}
\begin{split}
 n\geq n_{\mathrm{tens}}(r)4^{(R-r+1)c} 
 & = 4^2n_{\mathrm{tens}}(r)D[r+1]^c
 \end{split}
\end{equation}
for all $r$.  This allows us to simplify our expression for $N$ to get, with $K^{\prime}=4^{-2}K$,
\begin{equation}
\begin{split}
N   & \leq K^{\prime}n\sum_{r=R}^{1}D[r+1]^{n_\mathrm{mat}-c}\\
&\leq K^{\prime}n\left(1 + \sum_{r=R-1}^{1}D[r+1]^{n_\mathrm{mat}-c}\right)\\
&\leq K^{\prime}n \left(1 + \sum_{r=R-1}^{1}4^{(n_\mathrm{mat}-c)(R-r-1)}\right)\\
&\leq K^{\prime}n \left(1 +\sum_{x=0}^{R-2}4^{(n_\mathrm{mat}-c)x}\right)\\
& = K^{\prime}n \left(1 +\frac{1-4^{(n_\mathrm{mat}-c)(R-1)}}{1-4^{n_\mathrm{mat}-c}}\right)\\
& \leq K^{\prime}n \left(1 +\frac{4^{(n_\mathrm{mat}-c)(R-1)}}{4^{n_\mathrm{mat}-c}-1}\right)\\
& \leq K^{\prime}n \left(1 +\frac{4^{(n_\mathrm{mat}-c)R}}{4^{n_\mathrm{mat}-c}(4^{n_\mathrm{mat}-c}-1)}\right)\\
& \leq K^{\prime}n \left(1 +\frac{n^{(n_\mathrm{mat}-c)/c}}{4^{n_\mathrm{mat}-c}(4^{n_\mathrm{mat}-c}-1)}\right)\\
& \leq K^{\prime} \left(n +\frac{n^{n_\mathrm{mat}/c}}{4^{n_\mathrm{mat}-c}(4^{n_\mathrm{mat}-c}-1)}\right)\\
	& \leq bn^{\max(n_\mathrm{mat}/c,1)},
\end{split}
\end{equation}
where $b$ is a constant.  To get the second line, we used that $D[R+1] = 1$, and to get the eighth line, we used that $n\geq n_{\mathrm{tens}}(1)4^{Rc}=4^{Rc}$, which means that $4^R\leq n^{1/c}$.

A similar result holds for correlated noise provided it can be represented (or well approximated) by a matrix-product state on the boundary (physical) qubits, i.e., noise with short-range correlations, such as the factored noise considered in \cite{CF18}.  In that case, the formula for the bond dimension at each radius in equation (\ref{eq:bond_dim}) will be modified to give $D[r] = \kappa \times 4^{R-r}$, where $\kappa$ is the bond dimension of the matrix product state describing the noise. 

Note that the decoder also depends on the error syndrome $\vec{s}$ via the pure error, which can be calculated via the inverse syndrome former $F$, which is an $n\times(n-k)$ matrix, which gives $E(\vec{s}\,)=F\vec{s}$.  In the worst case scenario, the complexity of this will be $O(n^2)$.  However, for holographic codes, $F$ will be extremely sparse, making this calculation quicker.

\section{Threshold}
\label{app:threshold}
\begin{figure}[ht!]
\includegraphics[width=\columnwidth]{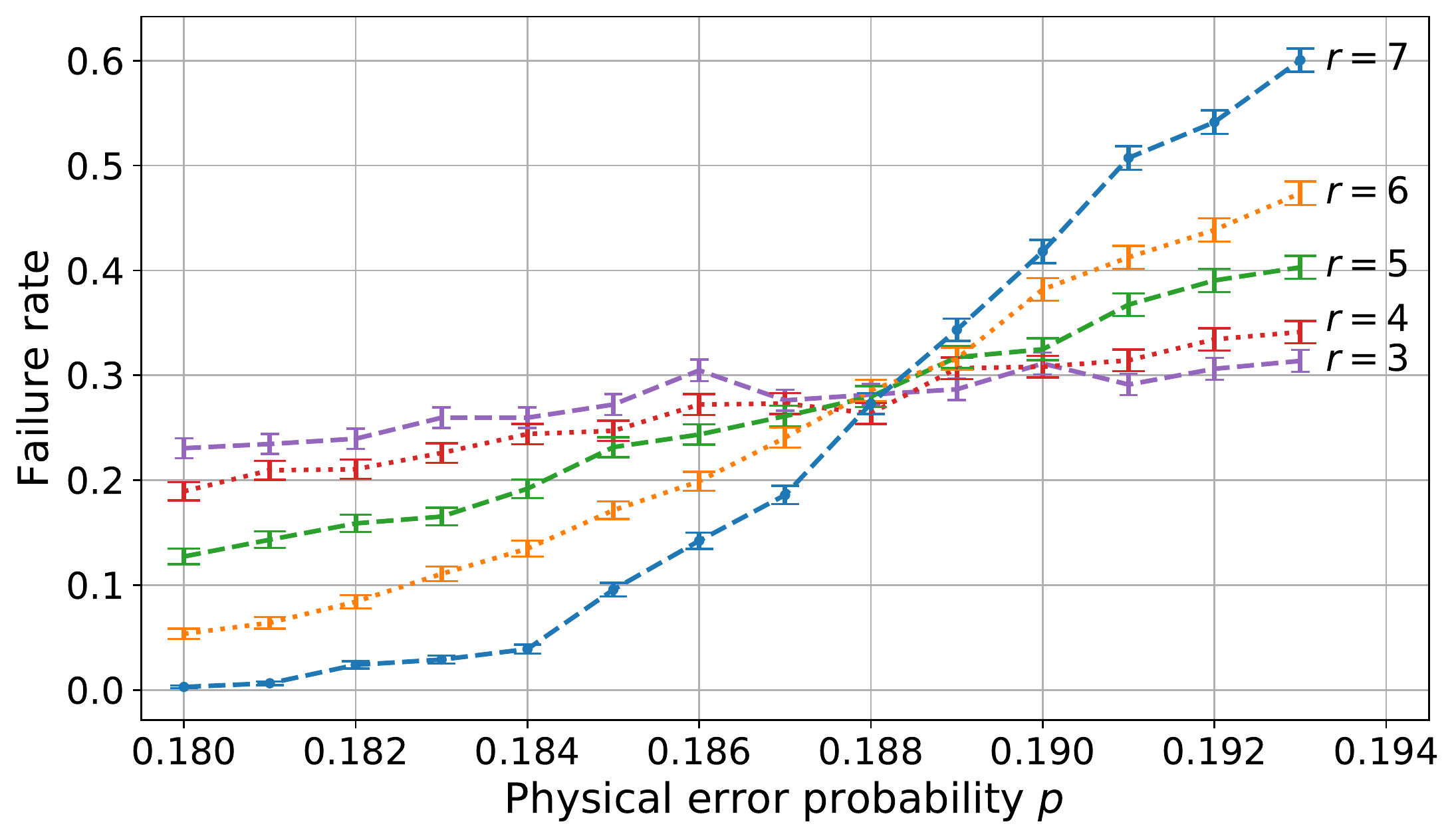}
\caption{Failure rate for the zero-rate six-qubit holographic code as a function of the single-qubit error probability (depolarizing noise parameter $p$) for different code radii.  Each point corresponds to $2000$ random errors, and the error bars correspond to standard errors.}
\label{fig:zoomed_thresh}
\end{figure}
Figure \ref{fig:zoomed_thresh} shows the failure probability versus the depolarizing noise strength $p$ close to the threshold.  Based on this, we would expect the threshold for the code $p_{\mathrm{th}}$ to be around $18.8\%$.  To confirm this, we can use a scaling hypothesis, analogously to \cite{WHP03}.  The idea is that, at the threshold, the success probability should be scale (i.e., code size) invariant.  And for any bigger error $p>p_{\mathrm{th}}$, larger codes will do worse, but for any smaller error $p<p_{\mathrm{th}}$, a larger code should do better.
\begin{figure}[ht!]
\includegraphics[width=\columnwidth]{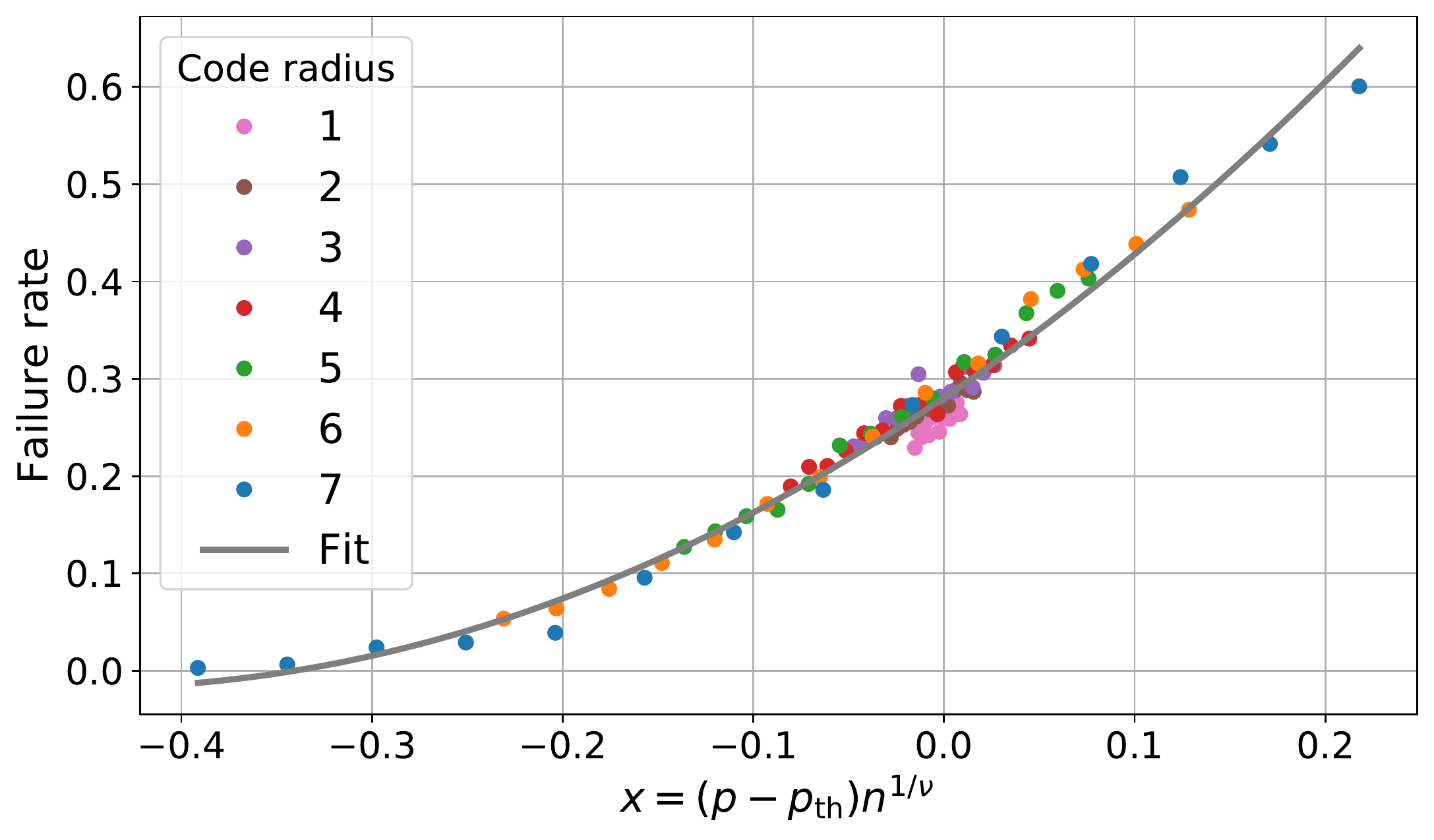}
\caption{Failure rate for the zero-rate six-qubit holographic code as a function of the rescaled error probability $x = (p-p_{\mathrm{th}})n^{1/\nu}$ for different code radii.  Using least-squares fitting, we fit a quadratic to the radius-seven data, as this is closest to the large-system limit.  Then we found the values of $\nu$ and $p_{\mathrm{th}}$ for which the data for all radii are as close to the curve as possible.  This yields values of $\nu = 2.970$ and $p_{\mathrm{th}}=0.188$.}
\label{fig:detailed_thresh}
\end{figure}

The next step is to think of there being an error correlation length $\xi$, which close to the critical point behaves like $\xi \sim |p-p_{\mathrm{th}}|^{-\nu}$, with critical exponent $\nu$.  The relevant parameter for us is rather $\xi/n$, where $n$ is the number of physical qubits.  This is because the physical qubits are a one-dimensional system (a circle at the boundary).  For errors to have a scale-invariant effect at the threshold, the error correlation length needs to grow with the system size.
So we should re-plot the failure probability as a function of $x=(p-p_{\mathrm{th}})n^{1/\nu}$.  The ansatz from \cite{WHP03} (for the surface code in their case) is that there is some universal function $f(x)$ for large codes describing the probability of failing to correct errors.  The function should satisfy $p_{\mathrm{failure}}\sim 3/4f(x)$ with $f(x)\rightarrow 1$ as $x\rightarrow \infty$ and $f(x)\rightarrow 0$ as $x\rightarrow -\infty$.

Our method was to fit a polynomial to the data for the largest code radius, giving the best approximation to $f(x)$.  Then we found which values of $\nu$ and $p_{\mathrm{th}}$ ensure that this estimate for $f(x)$ gives the best fit to the rest of the data.  This gave a value of $p_{\mathrm{th}}=0.188$ and $\nu=2.970$ as shown in figure \ref{fig:detailed_thresh}.

\end{document}